\documentclass[a4paper,11pt]{article}
\usepackage{pos}
\usepackage{cleveref}
\usepackage{amsmath,bm}
\usepackage{subcaption}
\usepackage{wrapfig}
\usepackage{enumitem}

\DeclareMathAlphabet{\mathcal}{OMS}{cmsy}{m}{n}

\title{Results on meson-meson scattering at large $\Nc$}

\author*[a,b]{Jorge Baeza-Ballesteros}
\author[b]{Pilar Hern\'andez,}
\author[c,d]{Fernando Romero-L\'opez}

\affiliation[a]{Deutsches Elektron-Synchroton DESY, Platanenallee 6, 15738 Zeuthen, Germany}
\affiliation[b]{IFIC, CSIC-Universitat de Val\`encia, 46980 Paterna, Spain}
\affiliation[c]{Albert Einstein Center, Institute for Theoretical Physics, University of Bern, 3012 Bern, Switzerland}
\affiliation[d]{Center for Theoretical Physics, Massachusetts Institute of Technology, USA}

\emailAdd{jorge.baeza.ballesteros@desy.de}
\emailAdd{m.pilar.hernandez@uv.es}
\emailAdd{fernando.romero-lopez@unibe.ch}

\abstract{We present results on the large $\Nc$ scaling of meson-meson scattering amplitudes. We work in a theory with $\Nf=4$ degenerate quark flavors and run lattice simulations with $\Nc=3-6$ and pion mass $\Mpi\approx 590$ MeV. We focus on three different scattering channels, two of which have the same quantum numbers as some tetraquark candidates recently found at LHCb. Finite-volume energies are extracted using a large set of operators, containing two-particle operators corresponding to two pions or two vector mesons, and local tetraquark operators. Using Lüscher's quantization condition, we constrain the infinite-volume scattering amplitudes and investigate subleading $\Nc$ corrections to the large $\Nc$ limit. For one of the channels, we find indications of a virtual bound state at $\Nc=3$, which may be related to one of the aforementioned exotic states. }

\FullConference{The 41st International Symposium on Lattice Field Theory (LATTICE2024)\\
 28 July - 3 August 2024\\
Liverpool, UK\\
Report numbers: DESY-25-018, MIT-CTP/5834\\}

\usepackage{cleveref}

\newcommand{\Nf}[0]{N_\text{f}}
\newcommand{\Nc}[0]{N_\text{c}}

\newcommand{\Mpi}[0]{M_\pi}
\newcommand{\Mrho}[0]{M_\rho}

\newcommand{\Fpi}[0]{F_\pi}

\newcommand{\cK}[0]{\mathcal{K}}
\newcommand{\cM}[0]{\mathcal{M}}
\newcommand{\cO}[0]{\mathcal{O}}
\newcommand{\cZ}[0]{\mathcal{Z}}

\newcommand{\LO}[0]{\mathrm{LO}}

\newcommand{\rcite}[1]{ref.~\cite{#1}}
\newcommand{\rrcite}[1]{refs.~\cite{#1}}

\begin{document}
\maketitle

\section{QCD in the large $\Nc$ limit}

The 't Hooft or large $\Nc$ limit of QCD~\cite{tHooft:1973alw} is the limit in which the number of colors, $\Nc$, is taken to infinity, while keeping the number of quark flavors, $\Nf$, constant. 
QCD simplifies in this limit, but retains most of its non-perturbative features, such as asymptotic freedom, confinement and spontaneous chiral symmetry breaking. The limit also makes it possible to characterize how different observables scale with $\Nc$ and $\Nf$ through a perturbative analysis of correlation functions at large $\Nc$. In many cases, large $\Nc$ predictions have proven accurate in the low-energy regime, and have been used in phenomenological approaches to QCD. In other cases, however, these predictions fail to reproduce experimental results, due to large subleading $\Nc$ corrections, which are hard to estimate analytically. The lattice regularization of QCD, on the other hand, enables the determination of these corrections from first principles.

Several works have studied the large $\Nc$ limit of QCD on the lattice---see \rcite{Hernandez:2020tbc} and references therein. We have recently investigated the scaling of different meson observables using lattice QCD simulations at varying $\Nc$ in a theory with $\Nf=4$ degenerate quark flavors.\footnote{All pseudoscalar mesons are degenerate in a theory with $\Nf=4$ degenerate quarks, and so we refer to them generically as ``pions'' ($\pi$).} These observables include the pion mass, $\Mpi$, and decay constant, $\Fpi$,~\cite{Hernandez:2019qed}, non-leptoning kaon decays~\cite{Donini:2020qfu}, leading to a better understanding of the $\Delta I=1/2$ puzzle, and pion-pion scattering near threshold~\cite{Baeza-Ballesteros:2022azb}---see also \rcite{DeGrand:2024lvp} for another recent study of the large $\Nc$ scaling of the pion-pion scattering length. 

In the context of meson-meson scattering, the large $\Nc$ limit has been used by different phenomenological investigations. A notorious example is \rcite{Pelaez:2006nj}, where the large $\Nc$ limit was used in combination to the inverse amplitude method (IAM)~\cite{Truong:1988zp,Dobado:1989qm} to investigate the dependence of resonance poles with $\Nc$. This study, however, neglects subleading $\Nc$ corrections. 

Another timely question is the possible existence of tetraquark states in the large $\Nc$ limit.  The standard lore, due to Witten~\cite{Witten:1979kh} and Coleman~\cite{Coleman:1980nk} precludes their existence based on the factorizations property. According to this property, the correlation function of a tetraquark state would decompose into that of two non-interacting mesons at large $\Nc$. Weinberg~\cite{Weinberg:2013cfa} recently presented counterarguments to this picture, indicating that tetraquarks could arise from poles in the subleading connected part of the correlation functions, and they would have a width $\Gamma_\text{tetra}\sim\cO(\Nc^{-1})$ analogously to other standard resonances. This latter claim was later refined in~\rrcite{Knecht:2013yqa,Cohen:2014tg}, finding that the width of tetraquark states would depend on their flavor content, with tetraquarks with four open flavors being narrower, $\Gamma_\text{tetra}\sim\cO(\Nc^{-2})$.

In this work, we study the $\Nc$ dependence of meson-meson scattering observables using lattice QCD simulations, extending the analysis in \rcite{Baeza-Ballesteros:2022azb} beyond threshold, with a two-fold objective. First, to characterize the $\Nc$ scaling of meson-meson scattering amplitudes, focusing on subleading corrections to the large $\Nc$ limit, and more specifically, to match lattice results to chiral perturbation theory (ChPT) and constrain the scaling of the relevant low-energy constants (LECs) from ChPT. Second, we aim to investigate the existence of tetraquark states in the limit and, if they are found, to characterize their nature as a function of $\Nc$.

We work in the same setup used in \rrcite{Hernandez:2019qed,Donini:2020qfu,Baeza-Ballesteros:2022azb}, with $\Nf=4$ degenerate quark flavors. In this context, meson-meson scattering observables classify in seven different scattering channels, corresponding to different irreducible representations (irreps) of the isospin group, SU(4)${}_\text{f}$, two of which are degenerate~\cite{Bijnens:2011fm}. We focus on three of them:
\begin{itemize}[itemsep=0.5pt,topsep=1pt,parsep=1pt]
\item The 84-dimensional irrep, known as the $SS$ channel, which is analogous to the isospin-two channel of two-flavor QCD.
\item The 20-dimensional irrep, known as the $AA$ channel, which is antisymmetric in both quarks and antiquarks, and only exists for $\Nf\geq 4$.
\item One of the two degenerate 45-dimensional irreps, known as the $AS$ channel, which is antisymmetric in quarks and symmetric in antiquarks.
\end{itemize}
Both the $SS$ and $AA$ channels only contain even partial waves, while the $AS$ channel contains odd partial waves. 

For all these channels, the large $\Nc$ limit allows one to characterize the $\Nc$ and $\Nf$ dependence of the scattering amplitudes. For the $SS$ and $AA$ channels, one finds
\begin{equation}\label{eq:SSAAlargeNc}
\cM^{SS,AA}_2=\mp\frac{1}{\Nc}\left(a+b\frac{\Nf}{\Nc}\pm c\frac{1}{\Nc}\right)+\cO(\Nc^{-3})\,.
\end{equation}
Both channels have a common large $\Nc$ limit with opposite sign, while subleading corrections are separated in two terms: one proportional to $\Nf$ that appears with the same sign, and another one independent of $\Nf$ with opposite signs. The coefficients $a$, $b$ and $c$ are numerical constants depending on the kinematics. In general, they cannot be determined analytically, but can be constrained using lattice simulations. What is more, by combining lattice results for both channels, it is possible to disentangle the two subleading terms even if simulating at fixed $\Nf=4$, as was done in \rcite{Baeza-Ballesteros:2022azb}. For the $AS$ channel, on the other hand,
\begin{equation}\label{eq:ASlargeNc}
\cM^{AS}_2=d\frac{\Nf}{\Nc^2}+ e\frac{1}{\Nc^2}+\cO(\Nc^{-3})\,.
\end{equation}
where $d$ and $e$ are again unknown constants. The scattering amplitude in this channel is suppressed by an additional power of $\Nc$, with the cancellation of the leading $\cO(\Nc^{-1})$ term being related to the kinematic properties of this channel.

In \rcite{Baeza-Ballesteros:2022azb}, both the $SS$ and $AA$ channels were investigated near threshold and results were matched to ChPT, successfully constraining subleading corrections to the large $\Nc$ scaling of the relevant LECs. The $AA$ channel was found to have attractive interactions, making it a candidate to contain a tetraquark state. A qualitative study based on the IAM supported the presence of such a state at higher center-of-mass (CM) energies. However, a quantitative study of this possibility requires of a dedicated lattice study, on which we focus in this work.

The putative existence of a tetraquark resonance in some of the channels under study is further supported by recent experimental findings. In recent years, LHCb has reported on the discovery of several scalar tetraquark states: the $T_{cs0}^0(2900)$ in the invariant mass of $D^-K^+$~\cite{LHCb:2020bls,LHCb:2020pxc}, and the  $T_{c\bar{s}0}^{++}(2900)$ and $T_{c\bar{s}0}^0(2900)$ in the invariant mass of $D_s^+\pi^+$ and $D_s^+\pi^-$, respectively~\cite{LHCb:2022sfr,LHCb:2022lzp}. While all these states are expected to be part of some isospin triplets in nature, they all lie on the $AA$ channel in a theory with $\Nf=4$ degenerate flavors. In addition, LHCb has a also found a vector tetraquark in the invariant mass of $D^- K^+$, known as $T_{cs1}^0(2900)$~\cite{LHCb:2020bls,LHCb:2020pxc}, which would lie on the $AS$ channel in our setup. It is worth remarking that all these exotic states have been phenomenologically described as vector-meson molecules, due to their proximity to the $D^* K^*$ and $D_s^* \rho$ thresholds~\cite{Molina:2022jcd}. 

\section{Lattice results for the finite-volume energy spectra}

Lattice simulations make it possible to determine two-particle finite-volume energies, which can be later used to constrain infinite-volume scattering observables~\cite{Luscher:1986pf,Luscher:1990ux}. In this work, we use four different ensembles with $\Nc=3-6$ at fixed $\Mpi\approx 590$ MeV and lattice spacing $a\approx 0.075$ fm, generated using the Iwasaki gauge action and $\Nf=4$ flavors of dynamical clover-improved Wilson fermions with periodic boundary conditions. Correlation functions are measured using a unitary setup with identical action on the valence sector. The lattices have spacial and temporal extent $(L/a)^3\times (T/a)=24^3\times 48$ ($\Nc=3$) and $20^3\times 36$ ($\Nc=4-6$). Simulations are performed using the HiRep code~\cite{DelDebbio:2008zf,DelDebbio:2009fd}.

To determine finite-volume energies, we use a large set of operators to compute a matrix of correlation functions. More concretely, we three types of operators: operators with the form of two-meson states with different relative momenta, either two pions ($\pi\pi$) or two vector mesons ($\rho\rho$), and local tetraquark operators. In all cases, we consider several values of the total momentum, $\bm{P}$, and study all cubic group irreps that contain pion-pion states with $s$-wave or $p$-wave, for the $SS$ and $AA$, and $AS$ channels, respectively. Operators with the form of two vector mesons are included to accurately determine the finite-volume energy spectrum below the four-pion inelastic threshold, since vector mesons are stable in our ensembles, $\Mrho\sim (1.7-2)\Mpi$. The use of vector-meson and tetraquark operators, moreover, could be very important to identify the presence of a tetraquark resonance~\cite{Dudek:2012xn}. To compute correlation functions involving a two-particle operator at source, we use time- and spin-diluted $\mathbb{Z}_2\times \mathbb{Z}_2$ stochastic sources. For those cases with a tetraquark operator at source, on the other hand, we make use of point sources located on a sparse lattice~\cite{Detmold:2019fbk},
\begin{equation}
T_\text{sp}(x)=\sum_{\bm{x}\in \Lambda_\text{S}} \text{e}^{-i\bm{Px}}\left[\bar{q}_{f_1}(x)\Gamma_1 q_{f_2}(x) \bar{q}_{f_3}(x)\Gamma_2 q_{f_4}(x)\right]\,,
\end{equation}
where $x=(\bm{x},t)$, $\Gamma_i$ are products of Dirac gamma matrices, flavor indices are used to project to the channels of interest and 
\begin{equation}
\Lambda_\text{S}(t)=\left\{a(t/a,s_1+dn_1,s_2+dn_2,s_3+dn_3)\,|\,\,n_i\in\mathbb{Z},0\leq n_i < (L/a)/d,0\leq s_i < d\right\}\,,
\end{equation}
is the sparse lattice, with $d=4$ and $s_i$ some offsets that are randomly chosen on all configurations and ensure we recover the correct correlation functions. 

After computing the matrix of correlation functions, we use a generalization of the shift-reweighting technique~\cite{Dudek:2012gj} to mitigate thermal effects related to the finite time extent of our lattices. We then solve a generalized eigenvalue problem (GEVP)~\cite{Michael:1982gb,Luscher:1990ck} and extract the finite-volume energies from one-state correlated fits to the eigenvalues of the GEVP. These fits are repeated over several fit ranges and the final results are determined from an average using weights based on the Akaike Information Criterium~\cite{Jay:2020jkz}. The eigenvectors of the GEVP are related to the overlap of the operators into the finite-volume states, and are used to identify those states that are predominantly two pions, on which we focus our amplitude study.

\begin{figure}[t!]
   \centering
\includegraphics[width=\textwidth,clip]{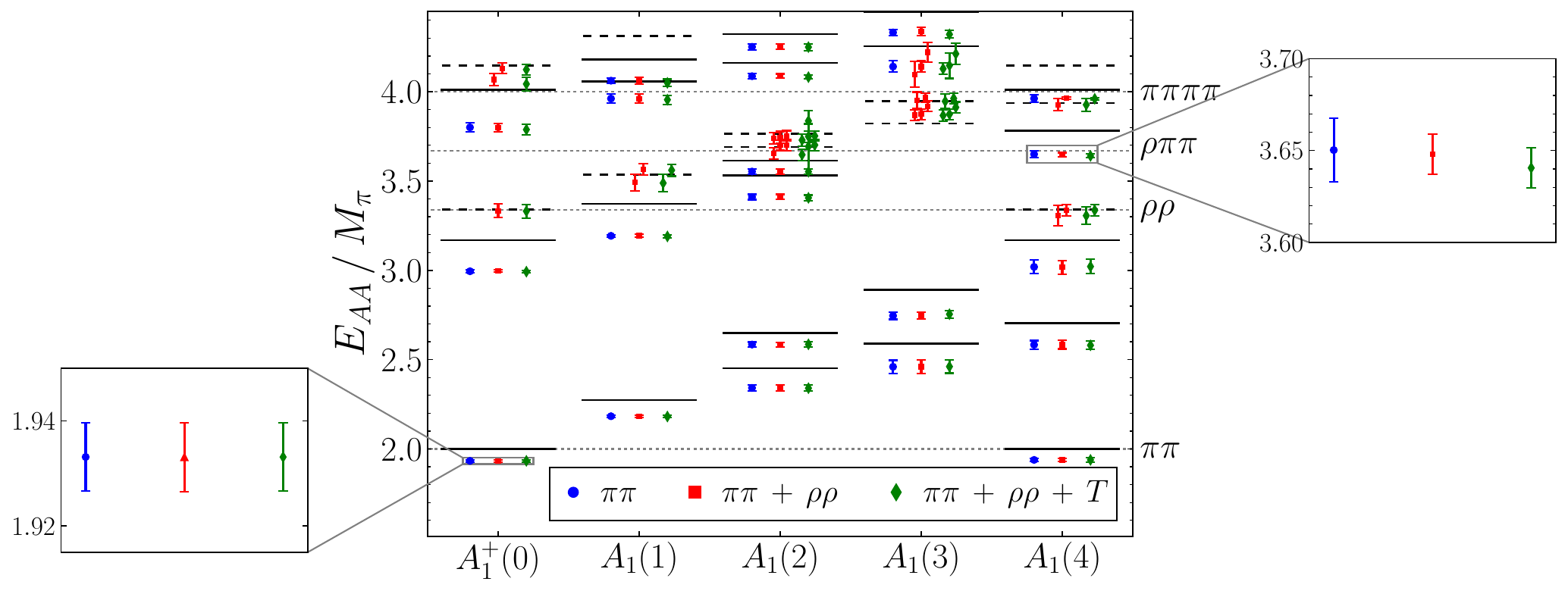}
   \caption{Preliminary results for the CM finite-volume energy spectra of the $AA$ channel with $\Nc=3$. Each column corresponds to a different irrep of the cubic group and momentum frame, with the number in parenthesis indicating $|\bm{P}|^2$ in units of $(2\pi/L)^2$, and each marker type refers to a different set of operators used to solve the GEVP. Horizontal segments are the free energies of two pions (solid) and two vector mesons (dashed), while dotted horizontal lines are the most relevant inelastic thresholds.}\label{fig:energiesvariation}
   
\end{figure}

It is instructive to study the impact on the results for the finite-volume energies of varying the operator set used to compute the matrix of correlation functions. In \cref{fig:energiesvariation} we show the finite-volume energy spectrum  for the $AA$ channel with $\Nc=3$, for three choices of the operator set. Horizontal solid and dashed lines are the free energies of two pions and two vector mesons, respectively, and dotted horizontal lines indicate relevant inelastic thresholds. We observe that the inclusion of vector-meson and tetraquark operators has little effect on the determination of the finite-volume energies of two-pion states, although it leads to the appearance of states that can be assigned to two vector mesons. The effect is completely negligible in the case of the lowest-lying states, although the inclusion of additional operators helps at reducing the errors for excited states. This can be observed in the zoomed-in panels in \cref{fig:energiesvariation}. 

\section{Results for the finite-volume scattering amplitude}

Two-particle finite-volume energies can be used to constrain infinite-volume scattering amplitudes using the so-called two-particle quantization condition (QC). This was first developed for two identical scalar particles~\cite{Luscher:1986pf,Luscher:1990ux}, and has since be extended to include any two-particle system~\cite{Luscher:1991cf,Rummukainen:1995vs,Kim:2005gf,He:2005ey,Bernard:2008ax,Luu:2011ep,Briceno:2012yi,Briceno:2014oea,Gockeler:2012yj}. In general, the two-particle QC takes the form of a matrix equation,
\begin{equation}\label{eq:quantizationcondition}
\text{det}\left[\cK_2^{-1}(E)+F(L,\bm{P};E)\right]=0\,,
\end{equation}
where $\cK_2$ is the two-particle $K$-matrix, related to the infinite volume scattering amplitude, and $F$ is a geometric factor that only depends on the lattice geometry and the total momentum, and contains power-law finite-volume effects. The solutions to this equation are the finite-volume energies.

When considering the simple case of single-channel scattering in the lowest partial wave $\ell$, \cref{eq:quantizationcondition} can be reduced to an algebraic equation that relates the finite-volume energies to the scattering phase shift, $\delta_\ell$. In the case of $s$-wave, it takes the form,
\begin{equation}\label{eq:schannelQC}
k\cot\delta_0=\frac{2}{\gamma L \pi^{1/2}}\cZ_{00}^{\bm{P}}\left(\frac{kL}{2\pi}\right)\,, 
\end{equation}
where $\gamma$ is the boost factor to the CM frame, $k$ is the magnitude of the relative momentum in the CM frame and $\cZ$ is the generalized zeta function~\cite{Luscher:1986pf,Luscher:1990ux,Rummukainen:1995vs}. Similar algebraic relations also hold for $p$-wave interactions, although the exact form depends on the cubic group irrep~\cite{Dudek:2012xn}.

\begin{figure}[t!]
   \centering
  \begin{subfigure}[t]{0.48\linewidth}
\centering%
\includegraphics[width=1\textwidth,clip]{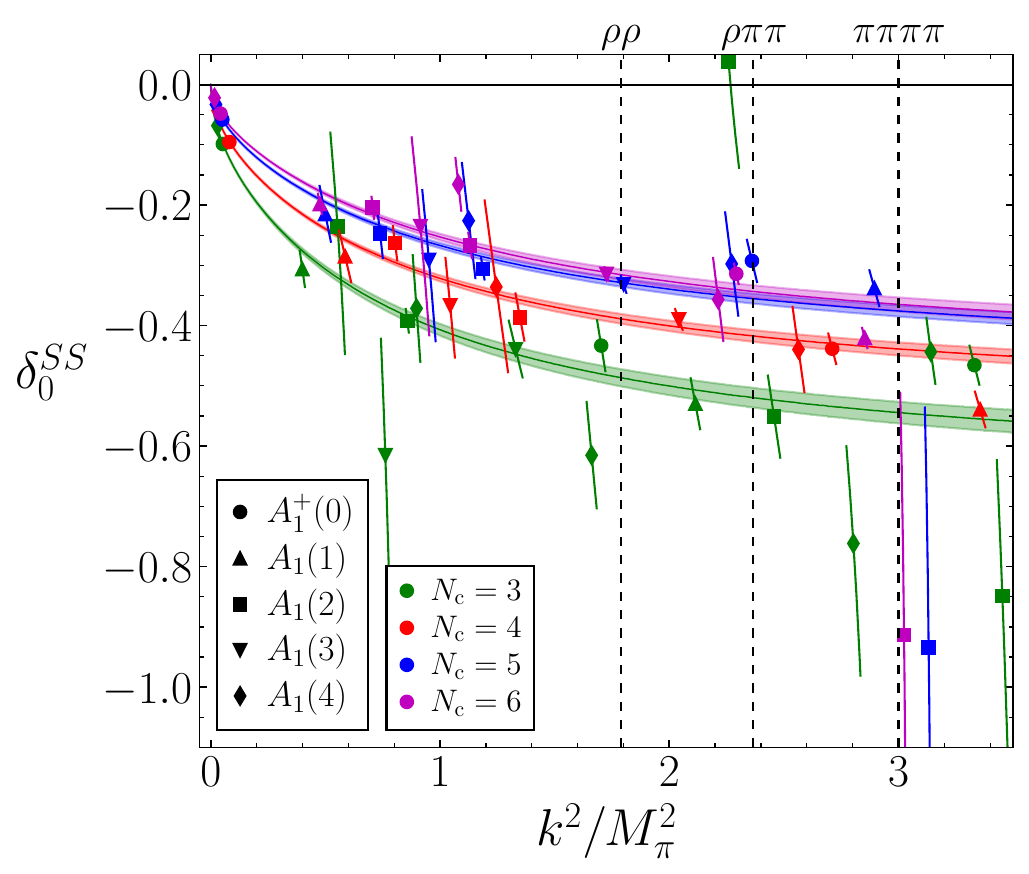}
\caption{$AA$ channel. }\label{fig:PSSS}
\end{subfigure}\hspace{0.02\textwidth}
   \begin{subfigure}[t]{0.48\linewidth}
\centering%
\includegraphics[width=1\textwidth,clip]{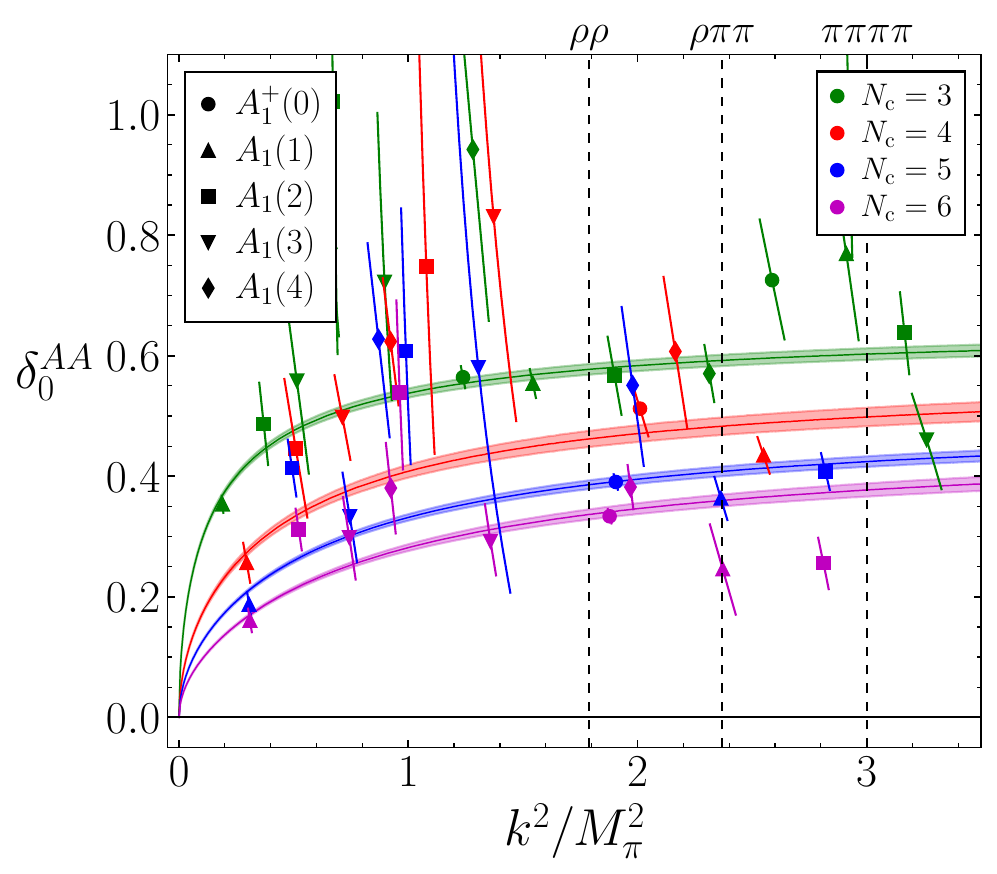}
\caption{$SS$ channel. }\label{fig:PSAA}
\end{subfigure}

   \caption{Preliminary results for the $s$-wave pion-pion scattering phase shift, for different $\Nc$. Dotted lines represent inelastic thresholds, as indicated over the figures, computed using the lightest value of $\Mrho$ among all ensembles. Solid lines and bands are the best-fit results to \cref{eq:modifiedERE} with $z=\Mpi^2$. }
   
\end{figure}

In \cref{fig:PSSS,fig:PSAA}, we present preliminary results for the pion-pion scattering phase shift in the $SS$ and $AA$ channels, respectively, obtained using \cref{eq:schannelQC}. We also indicate some of the most relevant inelastic thresholds, computed using the lightest mass of the vector and axial mesons among all ensembles. For both channels, we observe that the phase shift grows rapidly in magnitude above threshold, but depends weakly on the energy above $k\gtrsim \Mpi$. We also observe that the interactions become weaker as $\Nc$ increases, as expected from large $\Nc$ arguments.

To analyze the results in more detail, we constrain the scattering amplitude by fitting the finite-volume energies to predictions obtained with the QC for a given phase-shift parametrization. We perform fits to a modified effective range expansion~\cite{Yndurain:2002ud,Pelaez:2019eqa}, that includes the effect from an Adler zero,
\begin{equation}\label{eq:modifiedERE}
\frac{k}{\Mpi}\cot\delta_0=\frac{\Mpi E}{E^2-2z}\left(B_0+B_1\frac{k^2}{\Mpi^2}+...\right)\,,
\end{equation}
where we keep the location of the pole fixed at $z=\Mpi^2$ and set higher-order terms to zero. Fits are performed at fixed $\Nc$ and the results are presented as solid lines and one-sigma error bands in \cref{fig:PSSS,fig:PSAA}. 

In both channels, this parametrization leads to an accurate description of the lattice data. Our results for the $AA$ channel do not indicate the presence of any tetraquark resonance above threshold, but results  for $\Nc=3$ are consistent with a virtual bound state at $E_\text{b}/\Mpi=1.741(13)$. The nature of this state and its possible relation to experimental findings is under investigation.

\begin{figure}[t!]
   \centering
  \begin{subfigure}[t]{0.48\linewidth}
\centering%
\includegraphics[width=1\textwidth,clip]{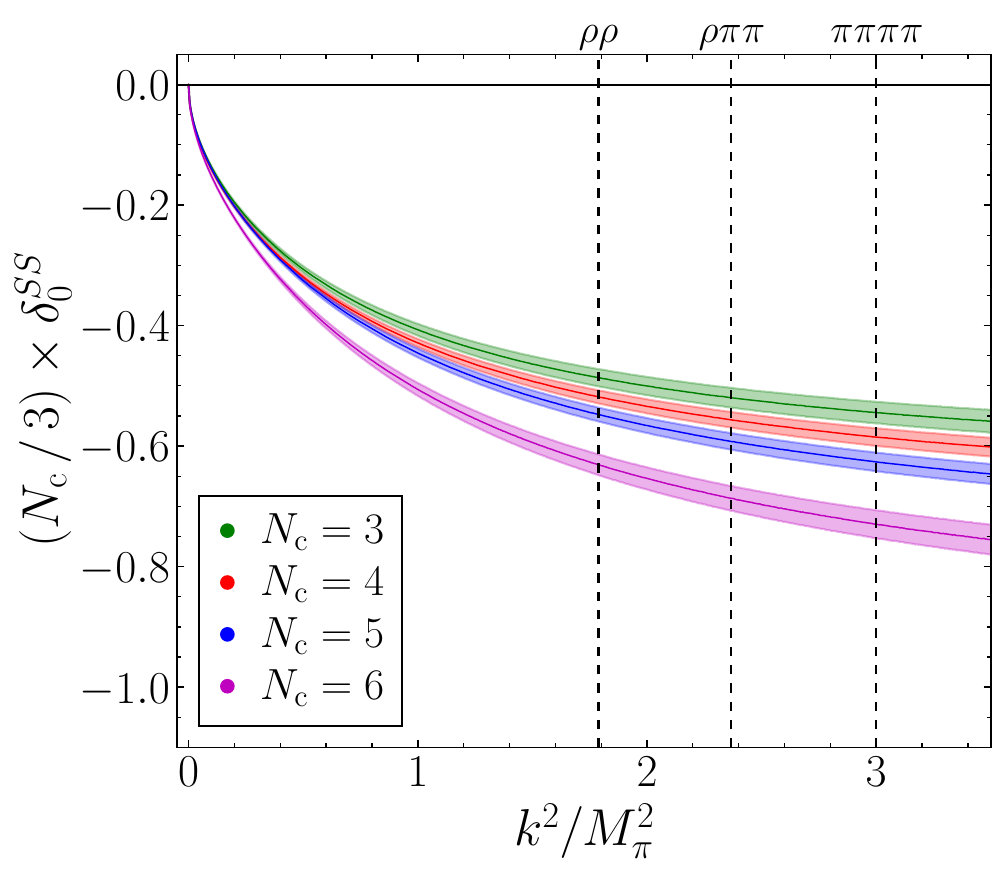}
\caption{$AA$ channel. }\label{fig:PSSSS}
\end{subfigure}\hspace{0.02\textwidth}
   \begin{subfigure}[t]{0.48\linewidth}
\centering%
\includegraphics[width=1\textwidth,clip]{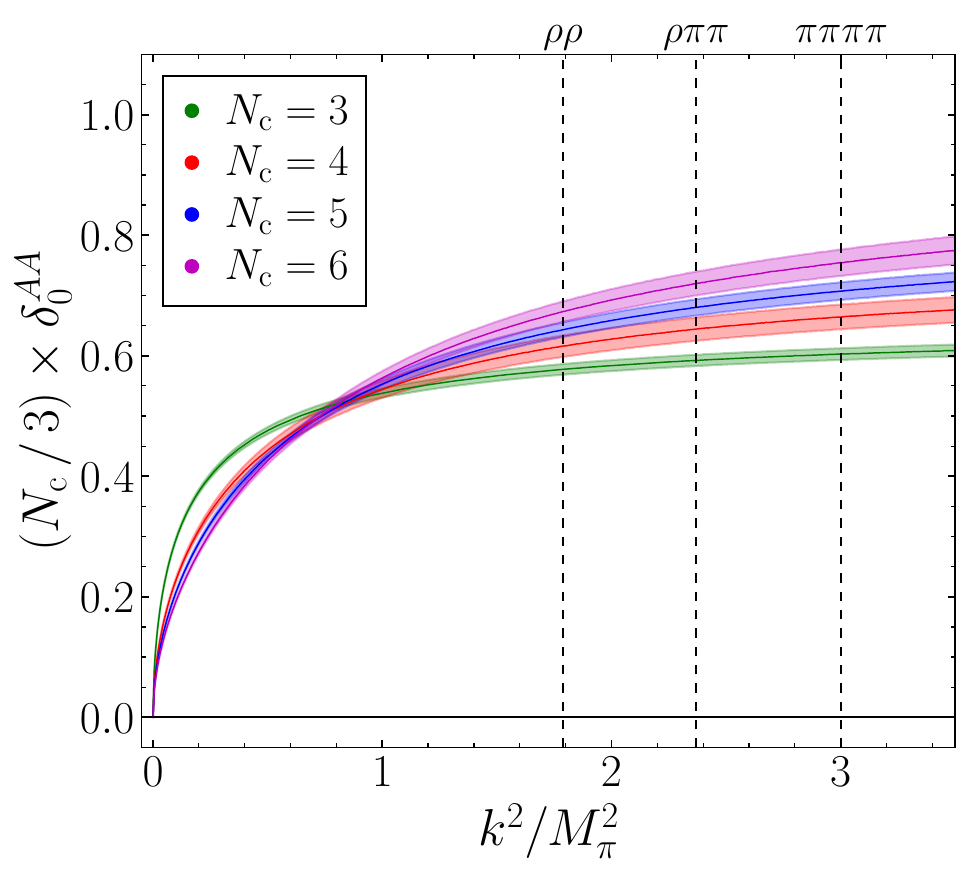}
\caption{$SS$ channel. }\label{fig:PSAAS}
\end{subfigure}

   \caption{Preliminary results for the $s$-wave pion-pion scattering phase shift, multiplied by $\Nc/3$ to eliminate the leading $\Nc$ dependence. }\label{fig:Ncsubleading}
   
\end{figure}

In addition, the fit results allow us to qualitatively study the sensitivity to subleading $\Nc$ corrections. In \cref{fig:Ncsubleading} we present these fit results multiplied by a factor that eliminates the expected leading $\Nc$ scaling. In both channels we observe that, after cancelling the leading $\Nc$ factors, the phase shift is of similar magnitude for all $\Nc$. However, we also note that we are sensitive to subleading corrections. In the $SS$ channel, results for $\Nc=6$ seem to be unnaturally separated from the rest, which we are currently investigating.

\begin{figure}[b!]
   \centering
  \begin{subfigure}[t]{0.48\linewidth}
\centering%
\includegraphics[width=1\textwidth,clip]{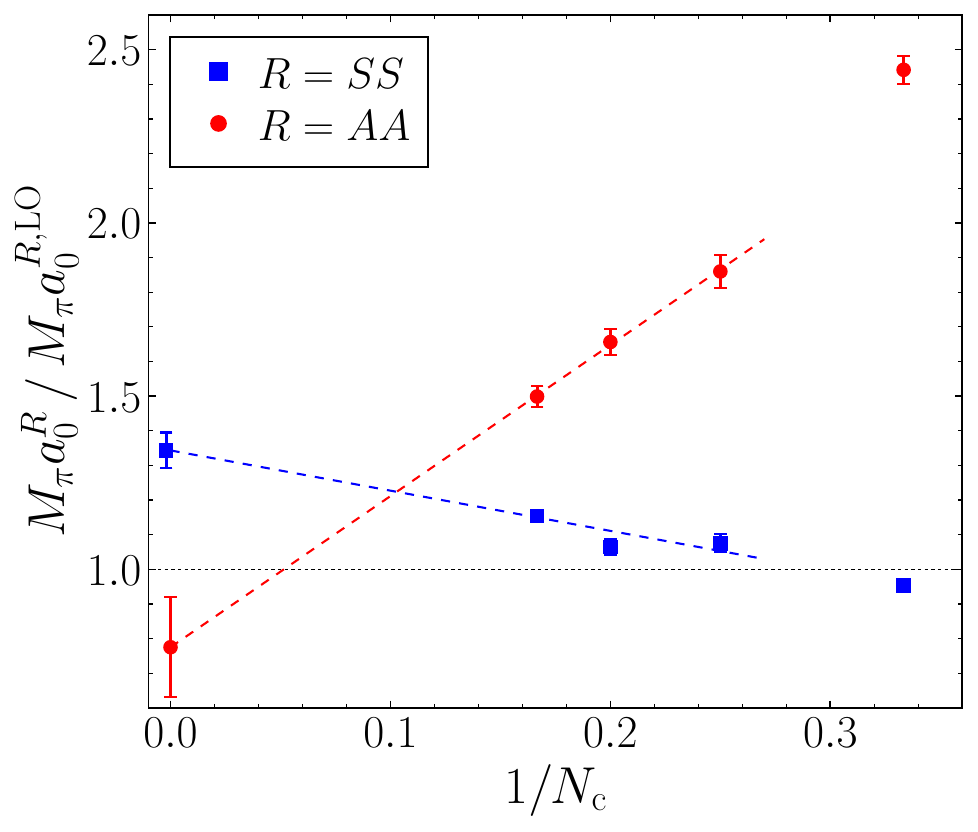}
\caption{Scattering length normalized by LO ChPT. }\label{fig:a0}
\end{subfigure}\hspace{0.02\textwidth}
   \begin{subfigure}[t]{0.48\linewidth}
\centering%
\includegraphics[width=0.995\textwidth,clip]{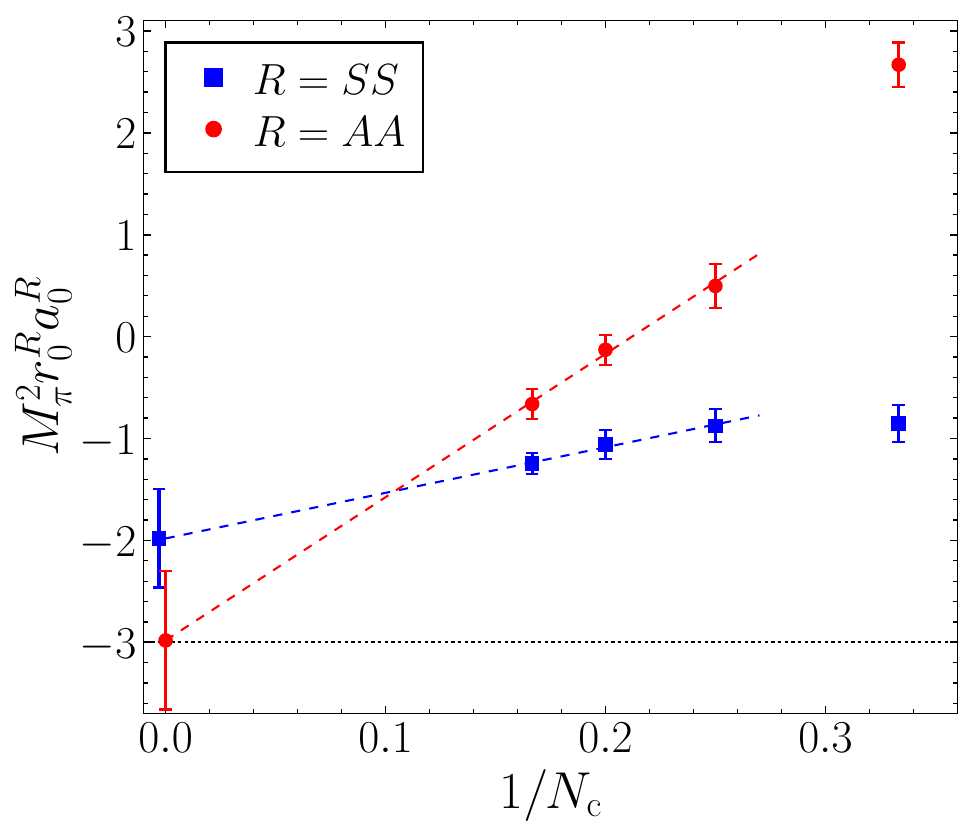}
\caption{Effective range. }\label{fig:r0}
\end{subfigure}

   \caption{Preliminary results for the large $\Nc$ dependence of the scattering length and effective range in the $SS$ and $AA$ channels, together with the LO predictions from ChPT. Results for $\Nc=4-6$ are used to linearly extrapolate to $\Nc\rightarrow \infty$. }\label{fig:lengthrange}
   
\end{figure}

We also analyze the sensitivity to subleading corrections by studying the $\Nc$ scaling of different scattering observables. In \cref{fig:lengthrange} we present results for the scattering length, $a_0$, and effective range, $r_0$, which can be related to the parameters in \cref{eq:modifiedERE}. Results for the scattering length are normalized by leading-order (LO) prediction from ChPT. For both quantities, we observe that our results for $\Nc=4-6$ are well described by a linear extrapolation towards the large $\Nc$ limit, with large $\cO(\Nc^{-2})$ corrections seemingly present for $\Nc=3$. However, our results for $a_0$ are not consistent between the two channels at large $\Nc$ as expected from \cref{eq:SSAAlargeNc}. This, however, could be related to mismatches in

\begin{wrapfigure}{r}{0.5\textwidth}
   \centering
   \begin{minipage}{0.47\textwidth}
   \includegraphics[width=1\textwidth,clip]{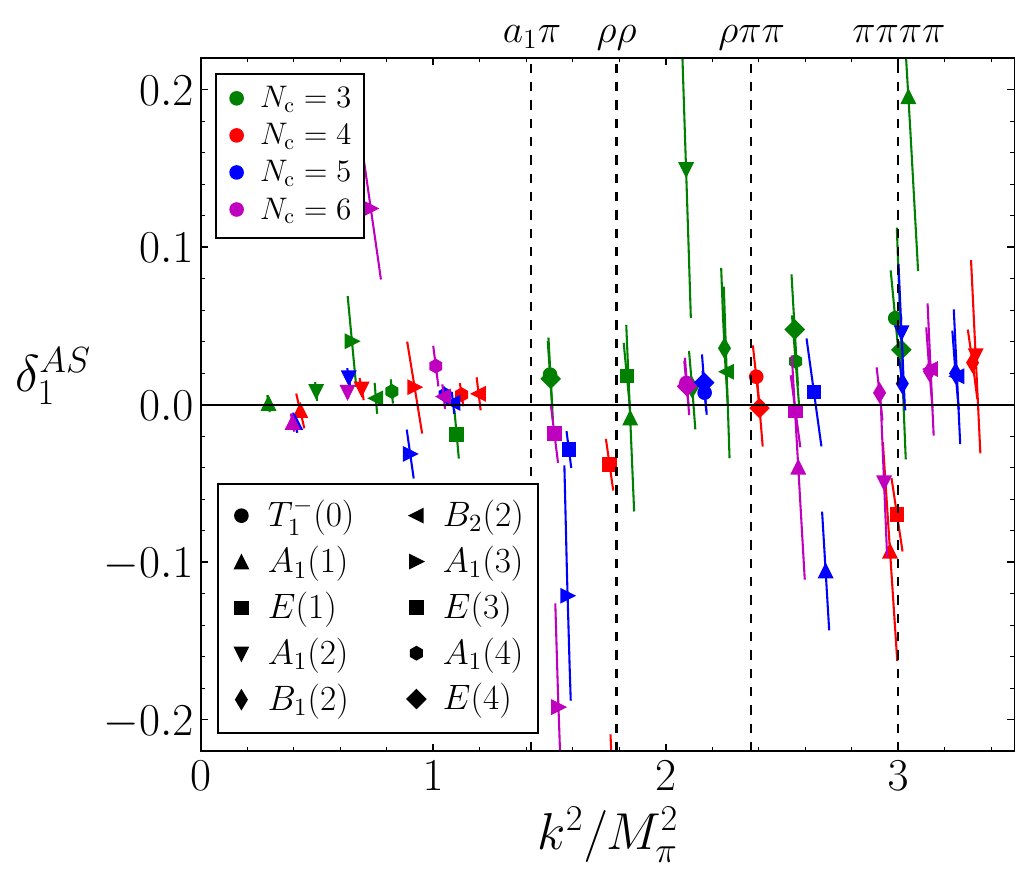}
   \caption{Preliminary results scattering phase shift in the $AS$ channel, for $\Nc=3-6$. }\label{fig:PSAS}
   \end{minipage}
\end{wrapfigure}

\noindent  the  pion masses \mbox{between} ensembles. To more accurately constrain the large $\Nc$ limit, one would need to incorporate the pion-mass dependence, which we plan to do via a fit to one-loop ChPT.

Finally, we present results for the $AS$ channel in \cref{fig:PSAS}, obtained using the $p$-channel counterparts of \cref{eq:schannelQC}. In this case, we observe the phase shift is very close to zero, indicating very weak interactions in this channel. This is in agreement with the large $\Nc$ limit, that predicts the scattering amplitude to be supressed as $\cO(\Nc^{-2})$---see \cref{eq:ASlargeNc}---, and also with ChPT, in which the amplitude is zero at LO, $\cM_2^{AS,\LO}=0$.

\section{Summary and outlook}

We have presented preliminary results for an ongoing study of the $\Nc$ dependence of meson-meson scattering amplitudes using lattice QCD, working in a theory with $\Nf=4$ degenerate quark flavors. Using a large set of operators, containing two-particle and local tetraquark operators, we have determined the finite-volume energy spectra for the $SS$, $AA$ and $AS$ channels with $\Nc=3-6$. These results have been used to constrain the infinite-volume pion-pion scattering phase shift up to the four-pion inelastic threshold. 

Preliminary results show sensitivity to subleading $\Nc$ corrections and we aim at matching these results to ChPT to constrain the $\Nc$ scaling of the LECs appearing in the effective theory. Our results also suggest the presence of a virtual bound state in the $AA$ channel for $\Nc=3$, which is currently under investigation. In particular, its existence needs to be established also with other parametrizations of the scattering amplitude. Should strong evidence of its existence be found, a compelling future venue could be the study of the dependence of the state on the pion mass, which might shed some light into the nature of the tetraquark states found at LHCb.

\acknowledgments

PH acknowledges support from the EU H2020 research and innovation programme under the MSC grant agreement No 860881-HIDDeN, and the Staff Exchange grant agreement No-101086085-ASYMMETRY, by the Spanish Ministerio de Ciencia e Innovaci\'on project PID2020-113644GB-I00, and by Generalitat Valenciana through the grant CIPROM/2022/69. The work of JBB has been partially supported by the aforementioned funding and the Spanish grant FPU19/04326 of MU. The work of FRL is supported in part by Simons Foundation grant 994314 (Simons Collaboration on Confinement and QCD Strings) and the Mauricio and Carlota Botton Fellowship.

\bibliographystyle{JHEP}
\bibliography{scattering.bib}

\providecommand{\href}[2]{#2}\begingroup\raggedright\begin{thebibliography}{10}

\bibitem{tHooft:1973alw}
G.~'t~Hooft, \emph{{A Planar Diagram Theory for Strong Interactions}},
  \href{https://doi.org/10.1016/0550-3213(74)90154-0}{\emph{Nucl. Phys. B}
  {\bfseries 72} (1974) 461}.

\bibitem{Hernandez:2020tbc}
P.~Hern\'andez and F.~Romero-L\'opez, \emph{{The large $N_{c}$ limit of QCD on
  the lattice}},
  \href{https://doi.org/10.1140/epja/s10050-021-00374-2}{\emph{Eur. Phys. J. A}
  {\bfseries 57} (2021) 52} [\href{https://arxiv.org/abs/2012.03331}{{\ttfamily
  2012.03331}}].

\bibitem{Hernandez:2019qed}
P.~Hern\'andez, C.~Pena and F.~Romero-L\'opez, \emph{{Large $N_c$ scaling of
  meson masses and decay constants}},
  \href{https://doi.org/10.1140/epjc/s10052-019-7395-y}{\emph{Eur. Phys. J. C}
  {\bfseries 79} (2019) 865}
  [\href{https://arxiv.org/abs/1907.11511}{{\ttfamily 1907.11511}}].

\bibitem{Donini:2020qfu}
A.~Donini, P.~Hern\'andez, C.~Pena and F.~Romero-L\'opez, \emph{{Dissecting the
  $\Delta I= 1/2$ rule at large $N_c$}},
  \href{https://doi.org/10.1140/epjc/s10052-020-8192-3}{\emph{Eur. Phys. J. C}
  {\bfseries 80} (2020) 638}
  [\href{https://arxiv.org/abs/2003.10293}{{\ttfamily 2003.10293}}].

\bibitem{Baeza-Ballesteros:2022azb}
J.~Baeza-Ballesteros, P.~Hern\'andez and F.~Romero-L\'opez, \emph{{A lattice
  study of \ensuremath{\pi}\ensuremath{\pi} scattering at large N$_{c}$}},
  \href{https://doi.org/10.1007/JHEP06(2022)049}{\emph{JHEP} {\bfseries 06}
  (2022) 049} [\href{https://arxiv.org/abs/2202.02291}{{\ttfamily
  2202.02291}}].

\bibitem{DeGrand:2024lvp}
T.~DeGrand, \emph{{Curve collapse for the isospin-2 pion scattering length from
  QCD with 3, 4, and 5 colors}},
  \href{https://arxiv.org/abs/2409.02242}{{\ttfamily 2409.02242}}.

\bibitem{Pelaez:2006nj}
J.R.~Pelaez and G.~Rios, \emph{{Nature of the f0(600) from its N(c) dependence
  at two loops in unitarized Chiral Perturbation Theory}},
  \href{https://doi.org/10.1103/PhysRevLett.97.242002}{\emph{Phys. Rev. Lett.}
  {\bfseries 97} (2006) 242002}
  [\href{https://arxiv.org/abs/hep-ph/0610397}{{\ttfamily hep-ph/0610397}}].

\bibitem{Truong:1988zp}
T.N.~Truong, \emph{{Chiral Perturbation Theory and Final State Theorem}},
  \href{https://doi.org/10.1103/PhysRevLett.61.2526}{\emph{Phys. Rev. Lett.}
  {\bfseries 61} (1988) 2526}.

\bibitem{Dobado:1989qm}
A.~Dobado, M.J.~Herrero and T.N.~Truong, \emph{{Unitarized Chiral Perturbation
  Theory for Elastic Pion-Pion Scattering}},
  \href{https://doi.org/10.1016/0370-2693(90)90109-J}{\emph{Phys. Lett. B}
  {\bfseries 235} (1990) 134}.

\bibitem{Witten:1979kh}
E.~Witten, \emph{{Baryons in the 1/n Expansion}},
  \href{https://doi.org/10.1016/0550-3213(79)90232-3}{\emph{Nucl. Phys. B}
  {\bfseries 160} (1979) 57}.

\bibitem{Coleman:1980nk}
S.R.~Coleman, \emph{{1/N}},  in \emph{{17th International School of Subnuclear
  Physics: Pointlike Structures Inside and Outside Hadrons}}, 3, 1980.

\bibitem{Weinberg:2013cfa}
S.~Weinberg, \emph{{Tetraquark Mesons in Large $N$ Quantum Chromodynamics}},
  \href{https://doi.org/10.1103/PhysRevLett.110.261601}{\emph{Phys. Rev. Lett.}
  {\bfseries 110} (2013) 261601}
  [\href{https://arxiv.org/abs/1303.0342}{{\ttfamily 1303.0342}}].

\bibitem{Knecht:2013yqa}
M.~Knecht and S.~Peris, \emph{{Narrow Tetraquarks at Large N}},
  \href{https://doi.org/10.1103/PhysRevD.88.036016}{\emph{Phys. Rev. D}
  {\bfseries 88} (2013) 036016}
  [\href{https://arxiv.org/abs/1307.1273}{{\ttfamily 1307.1273}}].

\bibitem{Cohen:2014tg}
T.D.~Cohen and R.F.~Lebed, \emph{{Are There Tetraquarks at Large $N_c$ in
  QCD(F)?}}, \href{https://doi.org/10.1103/PhysRevD.90.016001}{\emph{Phys. Rev.
  D} {\bfseries 90} (2014) 016001}
  [\href{https://arxiv.org/abs/1403.8090}{{\ttfamily 1403.8090}}].

\bibitem{Bijnens:2011fm}
J.~Bijnens and J.~Lu, \emph{{Meson-meson Scattering in QCD-like Theories}},
  \href{https://doi.org/10.1007/JHEP03(2011)028}{\emph{JHEP} {\bfseries 03}
  (2011) 028} [\href{https://arxiv.org/abs/1102.0172}{{\ttfamily 1102.0172}}].

\bibitem{LHCb:2020bls}
{\scshape LHCb} collaboration, \emph{{A model-independent study of resonant
  structure in $B^+\to D^+D^-K^+$ decays}},
  \href{https://doi.org/10.1103/PhysRevLett.125.242001}{\emph{Phys. Rev. Lett.}
  {\bfseries 125} (2020) 242001}
  [\href{https://arxiv.org/abs/2009.00025}{{\ttfamily 2009.00025}}].

\bibitem{LHCb:2020pxc}
{\scshape LHCb} collaboration, \emph{{Amplitude analysis of the $B^+\to
  D^+D^-K^+$ decay}},
  \href{https://doi.org/10.1103/PhysRevD.102.112003}{\emph{Phys. Rev. D}
  {\bfseries 102} (2020) 112003}
  [\href{https://arxiv.org/abs/2009.00026}{{\ttfamily 2009.00026}}].

\bibitem{LHCb:2022sfr}
{\scshape LHCb} collaboration, \emph{{First Observation of a Doubly Charged
  Tetraquark and Its Neutral Partner}},
  \href{https://doi.org/10.1103/PhysRevLett.131.041902}{\emph{Phys. Rev. Lett.}
  {\bfseries 131} (2023) 041902}
  [\href{https://arxiv.org/abs/2212.02716}{{\ttfamily 2212.02716}}].

\bibitem{LHCb:2022lzp}
{\scshape LHCb} collaboration, \emph{{Amplitude analysis of
  B0\textrightarrow{}D\textasciimacron{}0Ds+\ensuremath{\pi}- and
  B+\textrightarrow{}D-Ds+\ensuremath{\pi}+ decays}},
  \href{https://doi.org/10.1103/PhysRevD.108.012017}{\emph{Phys. Rev. D}
  {\bfseries 108} (2023) 012017}
  [\href{https://arxiv.org/abs/2212.02717}{{\ttfamily 2212.02717}}].

\bibitem{Molina:2022jcd}
R.~Molina and E.~Oset, \emph{{Tcs\textasciimacron{}(2900) as a threshold effect
  from the interaction of the D*K*, Ds*\ensuremath{\rho} channels}},
  \href{https://doi.org/10.1103/PhysRevD.107.056015}{\emph{Phys. Rev. D}
  {\bfseries 107} (2023) 056015}
  [\href{https://arxiv.org/abs/2211.01302}{{\ttfamily 2211.01302}}].

\bibitem{Luscher:1986pf}
M.~Luscher, \emph{{Volume Dependence of the Energy Spectrum in Massive Quantum
  Field Theories. 2. Scattering States}},
  \href{https://doi.org/10.1007/BF01211097}{\emph{Commun. Math. Phys.}
  {\bfseries 105} (1986) 153}.

\bibitem{Luscher:1990ux}
M.~Luscher, \emph{{Two particle states on a torus and their relation to the
  scattering matrix}},
  \href{https://doi.org/10.1016/0550-3213(91)90366-6}{\emph{Nucl. Phys. B}
  {\bfseries 354} (1991) 531}.

\bibitem{DelDebbio:2008zf}
L.~Del~Debbio, A.~Patella and C.~Pica, \emph{{Higher representations on the
  lattice: Numerical simulations. SU(2) with adjoint fermions}},
  \href{https://doi.org/10.1103/PhysRevD.81.094503}{\emph{Phys. Rev. D}
  {\bfseries 81} (2010) 094503}
  [\href{https://arxiv.org/abs/0805.2058}{{\ttfamily 0805.2058}}].

\bibitem{DelDebbio:2009fd}
L.~Del~Debbio, B.~Lucini, A.~Patella, C.~Pica and A.~Rago, \emph{{Conformal
  versus confining scenario in SU(2) with adjoint fermions}},
  \href{https://doi.org/10.1103/PhysRevD.80.074507}{\emph{Phys. Rev. D}
  {\bfseries 80} (2009) 074507}
  [\href{https://arxiv.org/abs/0907.3896}{{\ttfamily 0907.3896}}].

\bibitem{Dudek:2012xn}
{\scshape Hadron Spectrum} collaboration, \emph{{Energy dependence of the
  $\rho$ resonance in $\pi\pi$ elastic scattering from lattice QCD}},
  \href{https://doi.org/10.1103/PhysRevD.87.034505}{\emph{Phys. Rev. D}
  {\bfseries 87} (2013) 034505}
  [\href{https://arxiv.org/abs/1212.0830}{{\ttfamily 1212.0830}}].

\bibitem{Detmold:2019fbk}
W.~Detmold, D.J.~Murphy, A.V.~Pochinsky, M.J.~Savage, P.E.~Shanahan and
  M.L.~Wagman, \emph{{Sparsening algorithm for multihadron lattice QCD
  correlation functions}},
  \href{https://doi.org/10.1103/PhysRevD.104.034502}{\emph{Phys. Rev. D}
  {\bfseries 104} (2021) 034502}
  [\href{https://arxiv.org/abs/1908.07050}{{\ttfamily 1908.07050}}].

\bibitem{Dudek:2012gj}
J.J.~Dudek, R.G.~Edwards and C.E.~Thomas, \emph{{S and D-wave phase shifts in
  isospin-2 pi pi scattering from lattice QCD}},
  \href{https://doi.org/10.1103/PhysRevD.86.034031}{\emph{Phys. Rev. D}
  {\bfseries 86} (2012) 034031}
  [\href{https://arxiv.org/abs/1203.6041}{{\ttfamily 1203.6041}}].

\bibitem{Michael:1982gb}
C.~Michael and I.~Teasdale, \emph{{Extracting Glueball Masses From Lattice
  {QCD}}}, \href{https://doi.org/10.1016/0550-3213(83)90674-0}{\emph{Nucl.
  Phys. B} {\bfseries 215} (1983) 433}.

\bibitem{Luscher:1990ck}
M.~Luscher and U.~Wolff, \emph{{How to Calculate the Elastic Scattering Matrix
  in Two-dimensional Quantum Field Theories by Numerical Simulation}},
  \href{https://doi.org/10.1016/0550-3213(90)90540-T}{\emph{Nucl. Phys. B}
  {\bfseries 339} (1990) 222}.

\bibitem{Jay:2020jkz}
W.I.~Jay and E.T.~Neil, \emph{{Bayesian model averaging for analysis of lattice
  field theory results}},
  \href{https://doi.org/10.1103/PhysRevD.103.114502}{\emph{Phys. Rev. D}
  {\bfseries 103} (2021) 114502}
  [\href{https://arxiv.org/abs/2008.01069}{{\ttfamily 2008.01069}}].

\bibitem{Luscher:1991cf}
M.~Luscher, \emph{{Signatures of unstable particles in finite volume}},
  \href{https://doi.org/10.1016/0550-3213(91)90584-K}{\emph{Nucl. Phys. B}
  {\bfseries 364} (1991) 237}.

\bibitem{Rummukainen:1995vs}
K.~Rummukainen and S.A.~Gottlieb, \emph{{Resonance scattering phase shifts on a
  nonrest frame lattice}},
  \href{https://doi.org/10.1016/0550-3213(95)00313-H}{\emph{Nucl. Phys. B}
  {\bfseries 450} (1995) 397}
  [\href{https://arxiv.org/abs/hep-lat/9503028}{{\ttfamily hep-lat/9503028}}].

\bibitem{Kim:2005gf}
C.h.~Kim, C.T.~Sachrajda and S.R.~Sharpe, \emph{{Finite-volume effects for
  two-hadron states in moving frames}},
  \href{https://doi.org/10.1016/j.nuclphysb.2005.08.029}{\emph{Nucl. Phys. B}
  {\bfseries 727} (2005) 218}
  [\href{https://arxiv.org/abs/hep-lat/0507006}{{\ttfamily hep-lat/0507006}}].

\bibitem{He:2005ey}
S.~He, X.~Feng and C.~Liu, \emph{{Two particle states and the S-matrix elements
  in multi-channel scattering}},
  \href{https://doi.org/10.1088/1126-6708/2005/07/011}{\emph{JHEP} {\bfseries
  07} (2005) 011} [\href{https://arxiv.org/abs/hep-lat/0504019}{{\ttfamily
  hep-lat/0504019}}].

\bibitem{Bernard:2008ax}
V.~Bernard, M.~Lage, U.-G.~Meissner and A.~Rusetsky, \emph{{Resonance
  properties from the finite-volume energy spectrum}},
  \href{https://doi.org/10.1088/1126-6708/2008/08/024}{\emph{JHEP} {\bfseries
  08} (2008) 024} [\href{https://arxiv.org/abs/0806.4495}{{\ttfamily
  0806.4495}}].

\bibitem{Luu:2011ep}
T.~Luu and M.J.~Savage, \emph{{Extracting Scattering Phase-Shifts in Higher
  Partial-Waves from Lattice QCD Calculations}},
  \href{https://doi.org/10.1103/PhysRevD.83.114508}{\emph{Phys. Rev. D}
  {\bfseries 83} (2011) 114508}
  [\href{https://arxiv.org/abs/1101.3347}{{\ttfamily 1101.3347}}].

\bibitem{Briceno:2012yi}
R.A.~Briceno and Z.~Davoudi, \emph{{Moving multichannel systems in a finite
  volume with application to proton-proton fusion}},
  \href{https://doi.org/10.1103/PhysRevD.88.094507}{\emph{Phys. Rev. D}
  {\bfseries 88} (2013) 094507}
  [\href{https://arxiv.org/abs/1204.1110}{{\ttfamily 1204.1110}}].

\bibitem{Briceno:2014oea}
R.A.~Briceno, \emph{{Two-particle multichannel systems in a finite volume with
  arbitrary spin}},
  \href{https://doi.org/10.1103/PhysRevD.89.074507}{\emph{Phys. Rev. D}
  {\bfseries 89} (2014) 074507}
  [\href{https://arxiv.org/abs/1401.3312}{{\ttfamily 1401.3312}}].

\bibitem{Gockeler:2012yj}
M.~Gockeler, R.~Horsley, M.~Lage, U.G.~Meissner, P.E.L.~Rakow, A.~Rusetsky
  et~al., \emph{{Scattering phases for meson and baryon resonances on general
  moving-frame lattices}},
  \href{https://doi.org/10.1103/PhysRevD.86.094513}{\emph{Phys. Rev. D}
  {\bfseries 86} (2012) 094513}
  [\href{https://arxiv.org/abs/1206.4141}{{\ttfamily 1206.4141}}].

\bibitem{Yndurain:2002ud}
F.J.~Yndurain, \emph{{Low-energy pion physics}},
  \href{https://arxiv.org/abs/hep-ph/0212282}{{\ttfamily hep-ph/0212282}}.

\bibitem{Pelaez:2019eqa}
J.R.~Pelaez, A.~Rodas and J.~Ruiz De~Elvira, \emph{{Global parameterization of
  $\pi \pi $ scattering up to 2 ${\mathrm {\,GeV}}$}},
  \href{https://doi.org/10.1140/epjc/s10052-019-7509-6}{\emph{Eur. Phys. J. C}
  {\bfseries 79} (2019) 1008}
  [\href{https://arxiv.org/abs/1907.13162}{{\ttfamily 1907.13162}}].

\end{thebibliography}\endgroup

\end{document}